\documentclass[aps,prl,reprint,groupedaddress,floatfix]{revtex4-1}
\usepackage[pdftex]{graphicx} 
\usepackage{epstopdf}
\usepackage{amssymb} 
\usepackage{amsmath}

\let\baraccent=\= 
\renewcommand{\=}[1]{\stackrel{#1}{=}} 

\renewcommand\Re{\operatorname{Re}}

\begin{document}

\title{Transient absorption and photocurrent microscopy show hot electron supercollisions describe the rate-limiting relaxation step in graphene}
\author{Matt W. Graham, Su-Fei Shi, Zenghui Wang, Daniel C. Ralph, Jiwoong Park and Paul L. McEuen}
\affiliation{Laboratory for Atomic and Solid State Physics, Cornell University, Ithaca, NY 14853, USA}
\affiliation{Department of Chemistry and Chemical Biology, Cornell University}
\affiliation{Kavli Institute at Cornell for Nanoscale Science, Ithaca, NY 14853, USA}
\begin{abstract}

Using transient absorption (TA) microscopy as a hot electron thermometer we show disorder-assisted acoustic-phonon supercollisions (SCs) best describes the rate-limiting relaxation step in graphene over a wide range of lattice temperatures ($T_l=$5-300 K), Fermi energies ($E_F=\pm0.35$ eV), and optical probe energies ($\sim$ 0.3 - 1.1 eV). Comparison with simultaneously collected transient photocurrent, an independent hot electron thermometer, confirms the rate-limiting optical and electrical response in graphene are best described by the SC-heat dissipation rate model, $H=A(T^3_e- T^3_l)$. Our data further shows the electron cooling rate in substrate supported graphene is twice as fast as in suspended graphene sheets, consistent with SC-model prediction for disorder.


 
\end{abstract}

\maketitle

With high electron mobility and uniform spectral response spanning the far-IR to visible regions, graphene is an attractive material for next generation optoelectronic devices such as fast photodetectors, bolometers and plasmonic devices.\cite{Mueller2010,Lemme2011,Bonaccorso2010,yan2012,Blake2008,Fong2012} Graphene was originally predicted to have long (up to ns) hot electron and hole ($e-h$) lifetimes resulting from its unusually large optic phonon energies and vanishing density of states.\cite{Bistritzer2009,Kubakaddi2009}. However, time-resolved experiments show the actual $e-h$ relaxation time is orders of magnitude faster.\cite{Dawlaty2008,Winnerl2011,Breusing2009} The mechanism for fast energy dissipation in graphene has been the subject of considerable debate, with differing reports advocating either optical phonon\cite{Wang2010, Rana2011} or disorder-mediated acoustic phonon decay pathways.\cite{Song2011a,Graham2013,Betz2013}  Here we measure the electronic heat dissipation rate $H=C_edT_e/dt$ using both transient absorption (TA) and transient photocurrent (TPC) thermometry. In particular, we report TA measurements in graphene while varying the lattice temperature, Fermi energy, and optical probe energy. Our data confirms that acoustic phonons supercollisions (SCs) best describe the rate-limiting heat dissipation kinetics over the wide range of these parameters. 

\begin{figure}[htbp]
\includegraphics[height=4in]{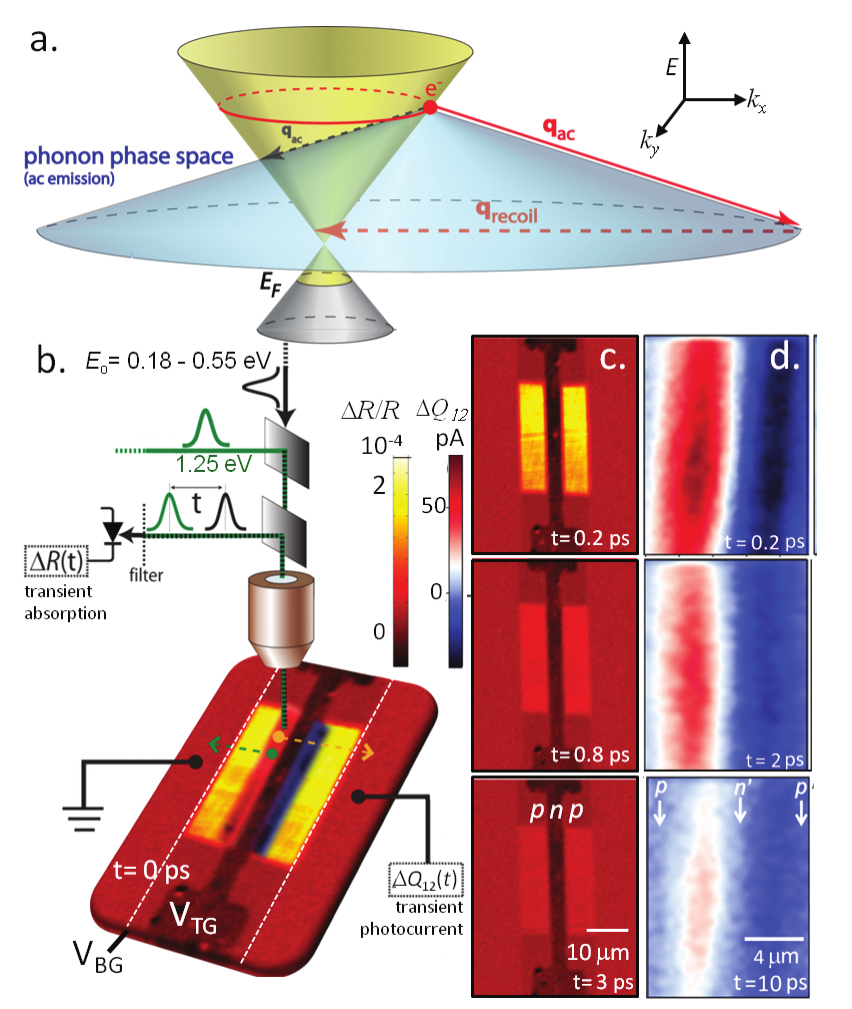}   
\caption{\textbf{Transient absorption + photocurrent}  \textbf{(a)} Supercollision cooling mechanism (\textit{red arrows}). \textbf{(b)} Measurement setup collects the optical TA, $\Delta R(t)$ and electrical TPC, $\Delta Q_{12}(t)f$ response from graphene (\textit{yellow}).  \textbf{(c)} Ultrafast TA movie frames for electron relaxation at $E_o=$0.4 eV probe, $T_l=5$ K. \textbf{(d)} TPC movie frames of electron relaxation at $p-n$ and $n-p$ graphene junctions, $T_l=5$ K. 
 }
\end{figure}  

In graphene, hot electrons can efficiently dissipate heat by emitting optical phonons with allowed energy, $\hbar \omega_{op} \sim $0.2 eV.\cite{Wang2010,Winnerl2011,Breusing2009} For electrons below this unusually high energy threshold, momentum conservation permits only low-energy ($<4$ meV, black arrows in Figs. 1a and 2a) acoustic phonon emission, resulting in very long electron relaxation times.\cite{Breusing2009,Bistritzer2009} However, Song \textit{et al.} predicts the SC model dominates where electron heat dissipation occurs without crystal momentum conservation, involving the emission of high-energy ($\sim k_B T_e$) acoustic phonons with the momentum imbalance, $q_{recoil}$ accounted for by disorder induced intrinsic lattice recoil.\cite{Song2011a}  This process, which results in faster cooling is depicted in Fig. 1a, and has a signature kinetic rate,\cite{Song2011a} 
\begin{align}
\frac{dT_e}{dt}=-\frac{H}{\alpha T_e}=-\frac{A}{\alpha}\frac{T_e^3-T_l^3}{T_e}.
\end{align}
where $A/\alpha$ is the SC rate coefficient, $T_l$ and $T_e$ are the lattice and electron temperatures, respectively. Solving Eq. 1, $T_e(t) \cong \frac{T_o}{1+AT_ot/\alpha}$ when $T_e(t) \gg T_l$ and $T_e(t) \cong T_l+(T_o-T_l)e^{-3AT_lt/\alpha}$ when $T_e(t)-T_l \ll T_l$ where $T_o$ is the initial electron temperature.  Recent studies demonstrate the SC-model\cite{Song2011a} successfully predicts graphene's photocurrent\cite{Graham2013} and electrical\cite{Betz2013} heating response. However, the applicability of the SC-model to purely optical measurements has not been considered.

In previous optical TA measurements hot electron cooling has instead been predominately modeled using the hot optical phonon (HP) cooling bottleneck effect.\cite{Wang2010,Huang2011} In the HP model, thermalized electrons (and holes) dissipate heat primarily by optic phonon emission from the Fermi-Dirac tails, where $E>\hbar \omega_{op}$. If electrons exchange their heat with optic phonons the two thermal baths are in approximate equilibrium, $T_{e}(t) \rightleftarrows T_{op}(t)$ over the lifetime, $\tau_{ph}$ of the dominate G-band optical phonon (see Fig. 2a\textsl{ii}).\cite{Wang2010} This forms a cooling bottleneck that determines the overall electronic temperature approximately given by $T_{e}(t)\cong T_{op}/(1+t/\tau_{op}),$ where $\tau_{op}^{-1}= k_BT_{op}/(\hbar \omega_{op}\tau_{ph})$ and $T_{op}$ is the initial optic phonon temperature.\cite{Wang2010}  In the $T_e(t) \gg T_l$ limit, both the HP and SC-models give identical functional forms. However, the two models make distinct predictions for the $T_l$ dependence, $E_F$ dependence and the role of environment-induced disorder.  
 
\begin{figure}[htbp]
\includegraphics[height=2.2in]{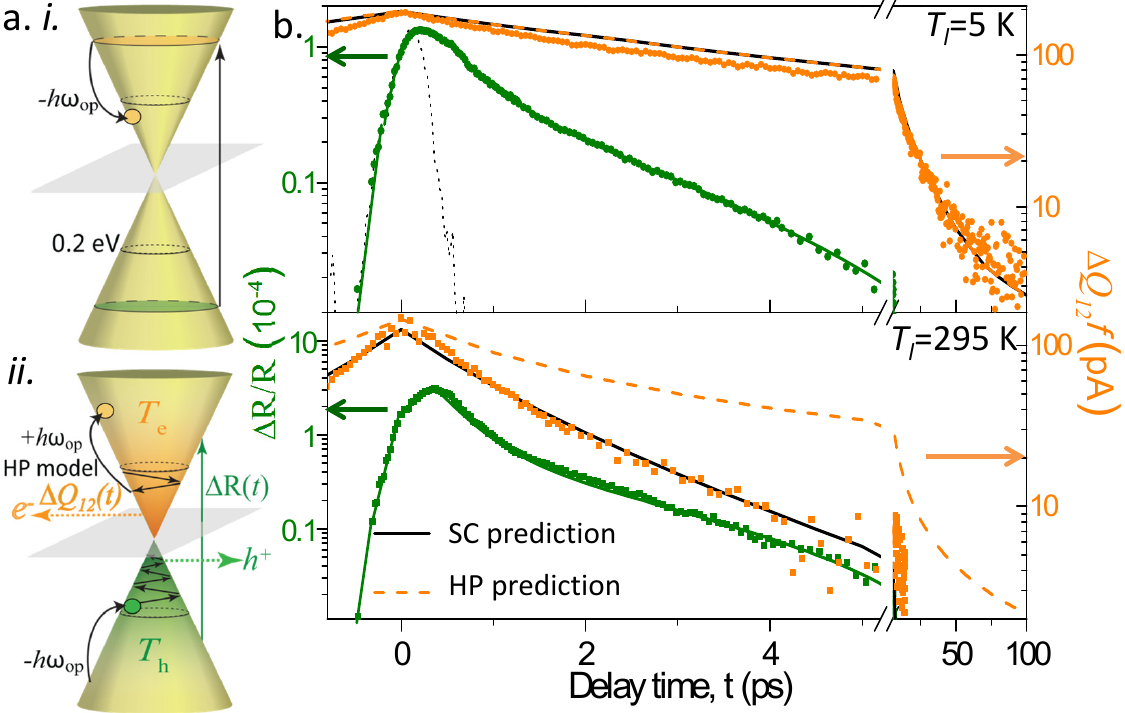}   
\caption{\textbf{SC vs. HP predictions} \textbf{(a)} Hot carriers thermalize, and emit optic or acoustic phonons (\textit{black arrows}).  We probe the electron temperature using TA and TPC. \textbf{(b)} Pulse cross-correlation (\textit{dotted line}).  The TA kinetics (\textit{green}) predict the TPC decay (\textit{orange}) when SC model is invoked (\textit{black lines}). The HP model (\textit{orange, dashed}) however, fails at 295 K.   
 }
\end{figure} 
 

To delineate between the HP and SC models, we compare each model against two independent thermometers of graphene temperature: (a) optical TA and (b) photothermal TPC. First, we optically probe the transient $e-h$ population of a single graphene sheet using confocal scanning TA microscopy.\cite{Huang2011} We detect both the spatial and temporal transient reflectivity, $\frac{\Delta R(t)}{R} = \frac{4}{n_s^2-1} \frac{4\pi}{c}\Re\Delta\sigma(E_{o},t)$, where $n_s$ is the substrate refractive index and $\Delta \sigma$ is the transient optical conductivity at energy, $E_o=\frac{1}{2}\hbar \omega_{probe}$.\cite{Malard2013} Fig. 1b shows our experiment setup where graphene at $T_l=$5 K is excited with a 170 fs pump pulse at 990 nm and the hot $e-h$ pairs created are probed at tunable wavelengths ranging from 1200 to 3450 nm.  In Fig. 1c, we observe a photobleach signal (\textit{yellow}) that corresponds to a $\sim 0.02$\% increase in probe beam reflectivity from Pauli blocking occurring at $E_o \pm E_F\cong0.4$ eV. The subsequent frames of this TA movie (\textit{see supplementary movie}) show hot electrons cooling uniformily. Figure 2b (\textit{green}) plots the kinetic decay of $\Delta R(t)/R$ at $T_l=5$ and 295 K obtained with a $1.5$ $\mu$m spot centered on a electrostatically doped graphene $p-n$ junction.  In this first TA measurement at a graphene $p-n$ junction, the kinetics exhibit a roughly $T_l$-independent biexponential decay similar to numerous existing graphene TA studies of single-layer graphene.\cite{Dawlaty2008,Breusing2009,Sun2008,Limmer2011}.


After electrons thermalize in graphene, the TA response is directly connected with a physical hot electron temperature, that is extracted by fitting to the transient interband optical conductivity, $\Delta \sigma(E_o,t)=-e^2/4\hbar \left[ f_{e/h}(T_e(t),E_o)-f_{e/h}(T_l,E_o)\right]$.  Absolute temperatures can be obtained by evaluating the Fermi-Dirac hot-electron occupancy probability, $f_{e/h}(T_e(t),E_o)$ at the energy ($E_o$) optically probing in the graphene band structure, giving approximately,\cite{Malard2013,CastroNeto2009,Kampfrath2005}  
\begin{equation}
\Delta \sigma(t) = \frac{\pi e^2}{2h} \left[ \tanh \left(\frac{E_o \pm E_F}{2k_BT_e(t)}\right)-\tanh \left( \frac{E_o \pm E_F}{2k_BT_l}\right) \right].
\end{equation}
We further show at our hot electron densities that the \textit{intra}band conductivity contributes negligibly to the transient reflectivity over our selected NIR probe regions (see supplemental section). Using the HP model, Eq. 2 predicts TA decays nearly exponentially.  The SC temperature model (Eq. 1) makes similar predictions only when $T_e \gg T_l$.  To fit the data in Fig. 2b, two exponents ($\tau_1$ and $\tau_2$) are required.  The faster component, $\tau_1 \cong 0.34$ ps averages over the initial electron thermalization and optic phonon emission timescale and is discussed elsewhere.\cite{Breusing2009,tielrooij2013}  Assuming the HP model describes the longer $\tau_2$ component, our fits to Fig. 2b with Eq. 2 requires $\tau_{ph}=2.9$ ps at 5 K and $\tau_{ph}=3.3$ ps at 295 K. 

If we instead apply the SC mechanism in Eq. 1, analytic fits to the TA response in Fig. 2b, yield rate coefficients of $A/\alpha$= 3.0$\times 10^{-4} K^{-1}ps^{-1}$ at 5 K and 4.4$\times 10^{-4} K^{-1}ps^{-1}$ and $T_o=1650\pm300$ K at 295 K.  A similar SC rate of $A/\alpha$= 5$\times 10^{-4} K^{-1}ps^{-1}$ was recently reported directly from PC measurements.\cite{Graham2013} This shows that the TA data can be explained using either the HP model or the SC model. However, the HP model predicts $\tau_{ph}$ values that are $>$2$\times$ longer than those measured via time-resolved Raman studies on near identical SiO$_2$ substrates.\cite{Kang2010,Wu2011}

We next use an independent thermometer to extract the hot electron temperature by simultaneously-collecting the graphene TPC response, shown in Figs. 1d and 2b(\textit{orange}).  Graphene's instantaneous photothermal current is given by $i(t)=\beta T_e(t)(T_e(t)- T_l)$, where $\beta$ is proportional to the Seebeck coefficient.\cite{Xu2009,Gabor2011,Graham2013} We detect the time-integrated current $Q_1f=f\int i(t,T_o) dt'$, where $f$ is the pulse repetition rate (76 MHz). After a delay time $t$, the electron gas cools and the second pulse at 0.8 eV re-heats graphene to a new initial temperature, $\sqrt{T_o^2+T_e(t)^2}$.  The TPC response, $Q_{12}$ is then obtained by integrating piecewise about $t$ giving,\cite{Graham2013}
\begin{equation}
Q_{12}(t)=\int^{t}_0{i(t',T_o)dt'}+\int_{t}^{\infty}{i(t'-t,\sqrt{T_o^2+T_e(t)^2})dt'}.
\end{equation}
In Fig. 1d we show the resulting PC autocorrelation function, $\Delta Q_{12}(t,r)=2Q_1-Q_{12}(t,r)$, decays in both time and space about the graphene $p-n$ junctions. In Fig. 2b, we plot t what our TA fit values ($\tau_{ph}$ for HP model and $A/\alpha$ for SC-model) predict for the TPC amplitude decay (\textit{dashed lines}).  While the HP model approximately predicts the $T_l=$5 K TPC response when $T_{op}$ is a free parameter, it clearly fails to predict rapidly decaying TPC response observed at room temperature (\textit{orange dashed lines}). On the contrary, the SC model correctly predicts the simultaneously acquired TPC kinetics at both $T_l=$5 and 295 K from their corresponding optical TA results.  


\begin{figure}[htbp]
\includegraphics[height=6in]{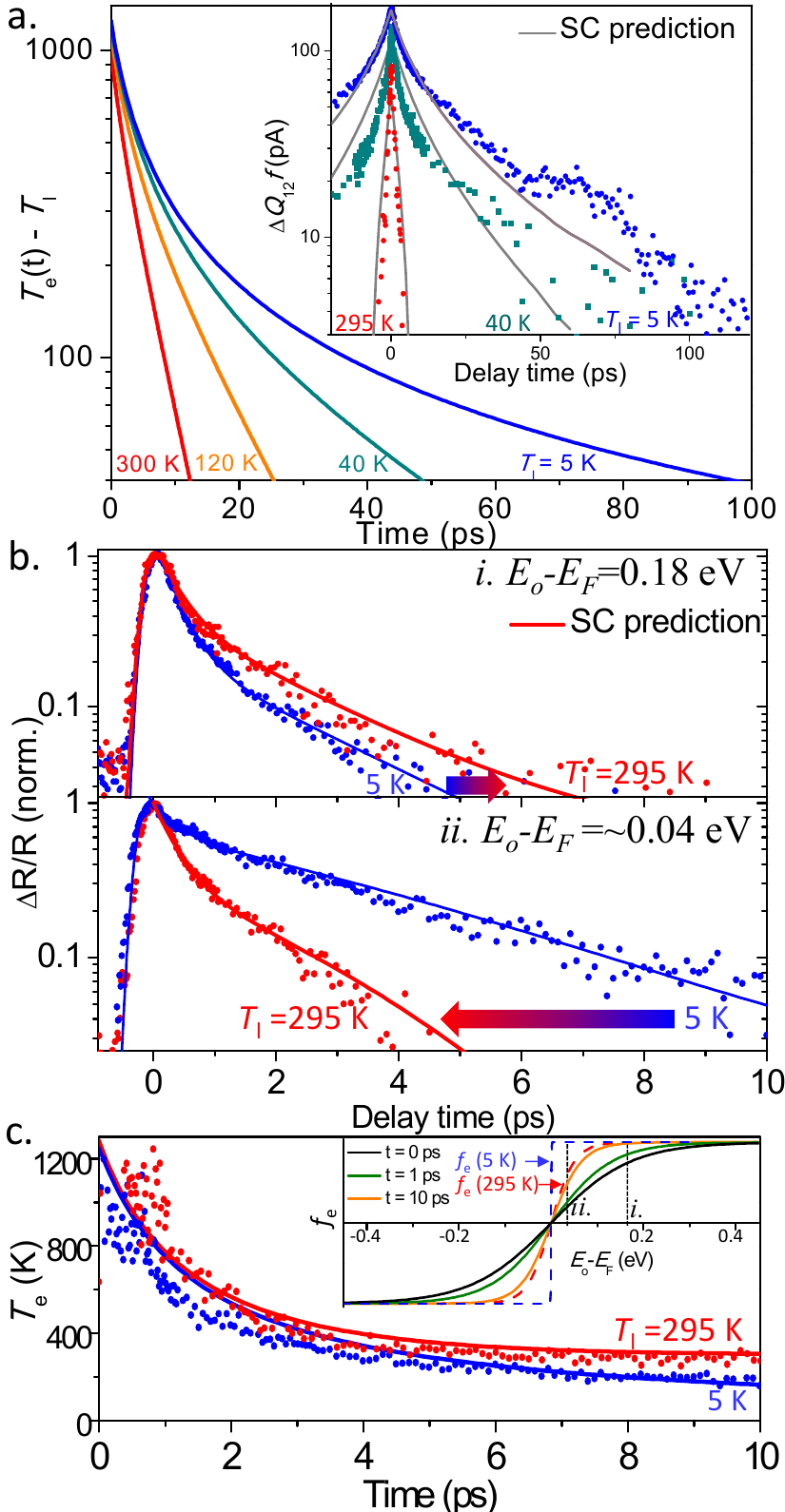} 
\caption{\textbf{Lattice temperature dependence.}  \textbf{(a)} Analytic SC-model solutions. (\textit{inset}) Using Eq. 3, the $T(t,T_l)$ curves predict the observed TPC decay with no free parameters.  \textbf{(b)} Mid-IR probe TA kinetics for (\textit{i}) $E_o=$0.18 eV and (\textit{ii}) $E_o-E_F\cong0.04$ eV. Using the 5 K TA parameters, the SC model predicts the 295 K result. \textbf{(c)} Solving Eq. 3 for $T_e$, we invert the data-points in 3b\textit{ii} and show $T_e(t)$ roughly agrees with the SC-model (\textit{solid lines}). (\textit{inset}) TA is proportional to the difference in the SC-model electronic occupancies, $f_e(T_e(t),E_o-E_F)$ (\textit{solid lines}) and $f_e(T_l,E_o-E_F)$ (\textit{dashed lines}). }
\label{fig4}
\end{figure}

The striking ability of the SC model to predict the electrical TPC kinetics from the optical TA  suggests both measurements can be well-described by the same SC heat loss rate, $H=A(T_e^3-T_l^3)$.  We now test the intrinsic $T_l$-dependence of the SC model independently for both TPC (Fig. 3a) and TA (Fig. 3b) measurements. With increasing $T_l$, the SC model solutions predict the cooling decay changes from a reciprocal to exponential decay in time (Fig. 3a).  Comparing this SC-model prediction against the TPC data in Fig. 3a(\textit{inset}), qualitative similarities are apparent. Here we numerically solve the TPC response function (Eq. 3) with no free parameters; as recently reported\cite{Graham2013}, we demonstrate again the SC model (\textit{gray lines}) predicts the TPC response.  Here, the TPC kinetics were acquired at pump fluences corresponding to $T_o \cong$1250 K at 295 K, and $\sim$850 K otherwise. For a fixed incident photon flux, the TPC decay is approximately independent of the excitation wavelength. 

Unlike TPC, in Fig. 2b our TA kinetic decay rate for relatively large $E_o$=0.4 eV, showed only a weak $T_l$ dependence. However, similar TA measurements performed with smaller $E_o$ ($\sim$0.18 eV) show markedly different behaviors (Fig. 3b).  Specifically, when $E_o-E_F<\hbar \omega_{op}$, strongly $T_l$ dependent kinetics emerge.  In Fig. 3b\textit{i}, we employ NIR-IR pump, mid-IR probe TA at $E_o=0.18$ eV $< \hbar \omega_{op}$, and plot the graphene mid-IR kinetics at $T_l=5$ K (\textit{blue}) and 295 K (\textit{red}). Fitting Fig. 3b\textit{i} at 5 K (\textit{blue line}) using Eqs. 1 and 2 plus a $\tau_1$ exponential component, we extract $\tau_1=0.36$ ps and $A/\alpha=2.4\times10^{-4}$ ps$^{-1}$K$^{-1}$. Using these 5 K TA parameters, we analytically solve the SC model at $T_l=$295 K, and show the result in Eq. 2 predicts the observed kinetics (\textit{red line}).  To probe closer to $E_F$, we next apply a back gate voltage such that $E_o-E_F\cong 0.04$ eV.  Fitting the remarkably longer $T_l=$5 K kinetics (\textit{blue line}) in Fig. 3b\textit{ii}, gives moderately faster SC-rate of $A/\alpha \cong 5.2\times10^{-4}$ ps$^{-1}$K$^{-1}$. Using this rate, we again solve the SC model to predict cooling at $T_l=$295 K, and the result closely predicts the radically faster kinetics observed (\textit{red line}).  We conclude the SC kinetic rate model predicts $T_l$-dependent TA response in graphene. 


The origin of the $T_l$-dependent TA can be understood by plotting the temporal evolution of the hot electron occupancy probabilities $f_e(T_e(t),E_o-E_F)$ in Fig. 3c(\textit{inset}, \textit{solid lines}). For a given $E_o-E_F$ probe window, $\Delta \sigma(t,E_o)$ is proportional to $f_e(T_e(t))- f_e(T_l)$, where $f_e(T_l)$ is the equilibrium electronic occupancy at $T_l=5$ and 295 K, respectively (\textsl{dashed lines}). At high probe energies, $f_e(295 K) \cong f_e(5 K) \cong 0$ making $\Delta \sigma(t,E_o)$ roughly $T_l$-independent, as observed in Figs. 2b and 3b\textit{i}. In contrast at low probe energies $f_e(295 K) \gg f_e(5 K)$, which makes hot electron kinetics effectively faster at room temperature as observed in Fig. 3b\textit{ii}.  Similar strong $T_l$-dependent responses have been reported in both recent THz studies\cite{Strait2011} and in degenerate far-IR TA measurements by Winnerl \textit{et al.}\cite{Winnerl2012,Limmer2011}   The SC-model roughly predicts these previously reported long lived transients, which were largely attributed to substrate heating effects before.

In Figs. 1-3, we demonstrated the SC model predicts the $T_l$-dependent TA interband electron kinetics across a wide range of probe energies.  Using the SC model, we may further invert our TA response to obtain $T_e(t)$.  In Fig. 3c we convert each data point in Fig. 3b\textit{ii} to its corresponding temperature, by solving Eq. 2 for $T_e(t)$, we also included a $\tau_1=0.35$ ps exponential component, accounting for non-thermalized electrons at short times. The resulting model-independent inversion approximately agrees with temperatures (\textit{solid lines}) obtained by directly solving Eq. 1, the SC-model.

So far, our SC-model predictions required prior knowledge of the intrinsic graphene doping to evaluate both initial temperature, $T_o$ and rate, $A/\alpha$. In Fig. 4 we study the $E_F$-dependence of the TA decay dynamics and extract two SC parameters: $T_o$ and $A/\alpha$. In Fig. 4a, we plot the $E_F$-dependence of the $T_l=5$ K mid-IR TA amplitude. By tuning $E_F$ via a capacitively coupled back-gate where $E_F \propto \sqrt{V_{BG}}$, we observe that both the TA signal amplitude (\textit{blue circles}) and lifetime (\textit{red squares}) increase as $E_o \rightarrow \pm E_F $. As previously observed, as $E_F > E_o$ the hot electron Pauli blocking effect is effectively turned off.\cite{Horng2011}  Accordingly, both the decay time and TA amplitude decrease in Fig. 4. The simple interband conductivity in Eq. 2 captures this overall trend well. The resulting fit (\textit{orange line}) requires an initial electron temperature of $\sim$1200 K, which agrees with estimates at similar fluences extracted earlier from TA SC-model fits or from photothermal current measurements.\cite{Graham2013}

In Fig. 4b, we systematically tune $E_F$ and plot the extracted 5 K rates $\tau_1^{-1}$ and $\tau_2^{-1}$ rates against $E_o-|E_F|$.  The initial TA decay rate is roughly invariant to $E_F$, with $\tau_1\cong0.36$ ps. Since this is longer than our 170 fs pulse width, this $E_F$ invariance  constant $\tau_1$ suggests that the electrons are not fully thermalized. When $T_e \gg T_l$, the SC model in Eq. 2 predicts $\tau_2^{-1}\cong (E_o \pm E_F) A/(k_B\alpha)$. Accordingly, we find the line of best fit intersects the origin with a slope of 2.6$\pm 0.1$ ps$^{-1}$eV$^{-1}$, which implies $A/\alpha=$2.3$\times 10^{-4} K^{-1}ps^{-1}$.  Instead of tuning $E_F$, we can systematically tune the probe energy $2E_o$, we find the extracted rate $\tau_2^{-1}$ also varies according to $(E_o \pm E_F) A/(k_B\alpha)$ at 5 K.  As shown in Fig. 5a (\textit{yellow squares}) the fitted slope gives $A/ \alpha$ = 2.3$\pm 0.4 \times 10^{-4}$ps$^{-1}$K$^{-1}$ and intercept gives $E_F= 190 \pm 90$ meV.  Thus, the 5 K TA kinetic dependence on both $E_F$ and $E_o$ give the same the same SC cooling rate.


\begin{figure}[htbp]
\includegraphics[height=3.5in]{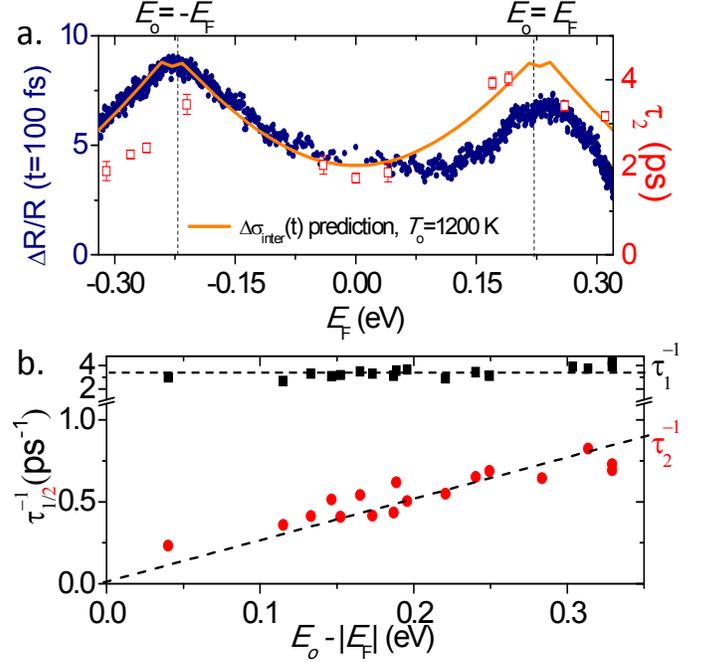}   
\caption{ \textbf{$\mathbf{E_F}$ dependence} \textbf{(a)} $E_o=0.18$ eV TA at $T_l= 5$ K vs. $E_F$.  As $\pm E_{F} \rightarrow \sim E_o$, both the transient amplitude and $\tau_2$ increase. The SC-model interband $\sigma(t)$ fits TA response well for $T_o=$1200 K. \textbf{(b)} TA lifetimes $\tau_2^{-1}$ scale linearly from the origin, with slope $A/\alpha k_B$. $\tau_1$ is constant at $\sim$0.36 ps. }
\end{figure}

\textit{Ab initio} predictions of the SC cooling rate are given in Song $et$ $al.$ as\cite{Song2011a,chen2012,Tse2009}: $\frac{A}{\alpha}=\frac{6 \zeta(3)}{\pi^2} \frac{\lambda}{k_F l} \frac{k_B}{\hbar} \cong \frac{2}{3}\frac{\lambda}{k_F l} \frac{k_B}{\hbar}$ where the electron-phonon coupling strength is $\lambda=\frac{D^2}{\rho s^2} \frac{2E_F}{\pi(\hbar v_F)^2}$.\cite{Song2011a} Using estimates for the deformation potential, $D = 10-30$ eV, $E_F$=0.1 eV and a mean free path of $k_F l =10$, this theory predicts: $A/\alpha= 10^{-4} - 10^{-3}$ K$^{-1}$ps$^{-1}$. (The range comes from the uncertainly in $D$). The best match to our experiments indicate $D = 8-14$ eV, well within the expected range. The SC model further predicts that $A/\alpha \propto E_F/k_Fl \propto  E_F/G $, where $G$ is the device conductivity.  For example over the back voltage sweep from 0 to 80 V, the $E_F$ changes from 0.1 to 0.3 eV, and our conductance changes from 0.1 to 0.4 mS; accordingly $A/\alpha$ vs. $E_F$ changes little. In rough accord with the SC-model,  in Fig. 4b $A/\alpha$ changes only from $2.7\times 10^{-4}$ ps$^{-1}$K$^{-1}$ to $2.0 \times 10^{-4}$ps$^{-1}$K$^{-1}$.

\begin{figure}[htbp]
\includegraphics[height=5in]{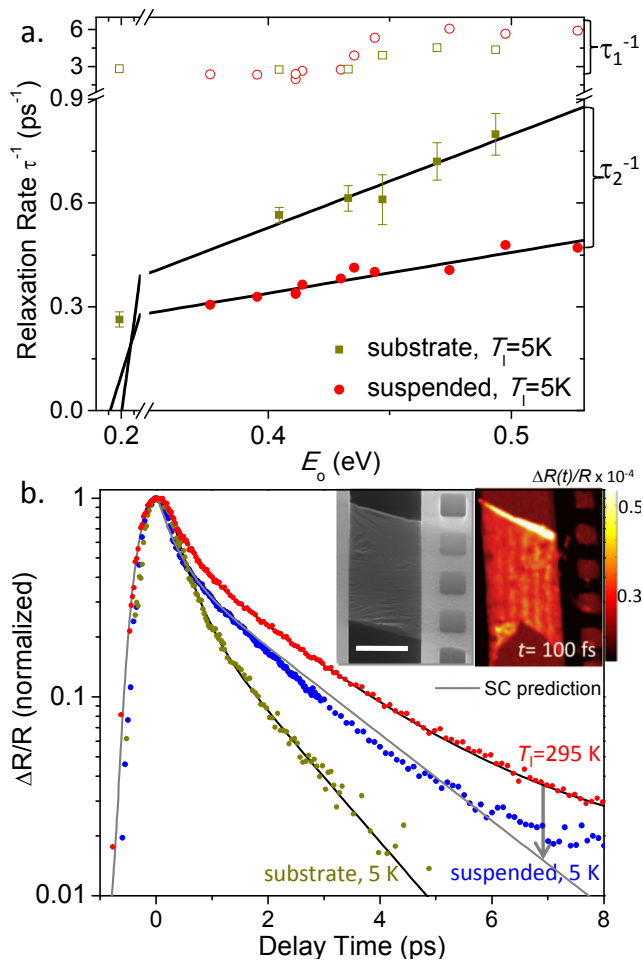}   
\caption{\textbf{Suspended graphene.}  \textbf{(a)} $\tau_2^{-1}$ varies linearly with probe energy, the intercept correspond to the intrinsic doping, $E_F$.  \textbf{(b)} Suspended TA kinetics for $E_o=0.4$ eV. The SC model predicts the 5 K decay with no free parameters (\textit{blue line}) from the 295 K TA parameters.  (\textit{inset}) SEM image and TA microscopy (at $t=0.1$ fs) of suspended graphene. The scale-bar is 10 $\mu$m. } 
\end{figure}

Short-range disorder is central to the SC-model, providing acoustic phonons with the requisite lattice recoil momentum ($q_{recoil}$, see Fig. 1a).\cite{Song2011a}  If we now suspend a graphene sheet in vacuum, how will this new environment impact the SC-cooling rate?  Fig. 5 shows the cooling mechanism in suspended graphene, resulting in cooling rate $\sim 2\times$ slower than its substrate supported counterpart. By plotting the decay rate $\tau_2^{-1}$ vs. probe energy in Fig. 5a, the linear fit line for suspended graphene (\textit{red circles}) requires $A/\alpha$ = 1.1$\pm0.2 \times 10^{-4}$ps$^{-1}$K$^{-1}$ and the intercept gives intrinsic doping at $E_F =95 \pm 32$ meV. Fitting suspended graphene kinetics using the alternate HP model would require a prohibitively long $\tau_{ph}=5.1$ ps optical phonon lifetime.\cite{Kang2010,Wu2011}


Fig. 5b compares the TA kinetics at $T_l=$5 K and 295 K.  Fitting the 295 K the decay using Eqs. 1-2, we again extract a roughly 2$\times$ slower rate coefficient, $A/ \alpha$ = 1.4$\pm0.1 \times 10^{-4}$ps$^{-1}$K$^{-1}$ and $T_o=850$ K.  The twofold slower cooling rate of suspended graphene vs. substrate supported is justified by the SC-prediction that $A/\alpha \propto (k_Fl)^{-1}$. Accordingly, transport studies have shown $k_Fl$ is approximately twice as long in suspended CVD-grown graphene vs. substrate suspended.\cite{Du2008,Bolotin2008,Adam}  Lastly we demonstrate the SC model predicts $T_l$-dependent hot electrons kinetics for suspended graphene, as demonstrated earlier for substrate-supported graphene.  Using the 295 K $A/\alpha$ extracted in Fig. 5b, we solve Eqs. 1 and 2 for the 5 K result.  We find the SC model predicts (\textit{gray line}) the 5 K suspended graphene TA kinetic decay with no free parameters.  



The ability of the SC model to predict graphene's optical, photocurrent,\cite{Graham2013} and electrical response\cite{Betz2013} under a wide variety of conditions definitively show the SC model best describes the rate-limiting heat dissipation step in photoexcited doped graphene.  Existing models, such as the HP model do not account for the strongly $T_l$-dependent kinetics observed in both TPC and TA measurements. The HP model further requires prohibitively long optic phonon lifetimes, $\tau_{ph} \sim$5 ps to fit suspended graphene kinetics. The SC-model provides a new interpretation for previous graphene TA studies, and suggests a new time line of event for electronic relaxation in graphene.  Over an initial timescale $\tau_1<0.4$ ps, photoexcited carriers rapidly thermalize and dissipate energy to optical phonons.\cite{Wang2010, Breusing2009} The vast majority of electrons now have $E<\hbar \omega_{op}$, resulting in a cooling bottleneck.  Here we show this bottleneck cools according to the SC-cooling kinetic rate law, $H_{SC}=A(T_e^3-T_l^3)$ with a rate coefficient determined by intrinsic disorder.  Collectively, joule heating \cite{Betz2013}, photocurrent \cite{Graham2013} and optical measurements can be described by the same SC-model.\cite{Song2011a} This suggests a reliable method for determining the electronic temperature in graphene has emerged.

\textbf{Experimental Methods:} 

The CVD growth, fabrication and characterization of both suspended and of $p-n$ junction graphene is found in the supplementary section. In $p-n$ junctions, a tunable back gate (BG) and top gate (TG) couple to graphene, defining two $p-n$ doped regions where the PC production is maximal. The collected PC amplitude is plotted as the laser is raster scanned over the $p-n$ junction (see superimposed PC map in Fig. 1a).  We optically excite the graphene $p-n$ junction region with pulses produced by two synchronously locked independently-tunable oscillators and NIR optical parametric oscillator (OPO).  Similar results were obtained when oscillator and white-light supercontinuum geometry is used.  We simultaneously collect the change in reflectivity ($\Delta R(t)/R$, TA) and electrical current generated ($\Delta Q_{12}(t)f$, TPC) as functions of pulse delay time. 

Cross-correlation at the device position yielding a 170 fs FWHM pulse duration.  After a mechanical delay stage, the two beams are aligned in a collinear geometry at and beamsplitter and coupled into the microscope (Olympus BX-51) through a 50XIR Olympus by a scanning mirror (SM, PI$\# S-334.2SL$). Mid-IR TA used $\sim3.5$ $\mu$m spot-size and reflective objective. TA signal was detected by lock-in detection at 0.9 MHz pump-beam AOM modulation rate using either amplified InGaAs or PbSe detectors. Pump power typically corresponds to an initial electron carrier density of $\sim 3\times 10^{12}$ cm$^{-2}$. Probe power was $\sim$1/20 of pump, unless specified. TPC was collected at 3 kHz modulation rate.

\begin{acknowledgments}
\textbf{Acknowledgments}:  research was supported by the Kavli Institute at Cornell for Nanoscale Science (KIC), AFOSR (FA 9550-10-1-0410), by the NSF through the Center for Nanoscale Systems and by the MARCO Focused Research Center on Materials, Structures, and Devices.  Device fabrication was performed at the Cornell Nanofabrication Facility/National Nanofabrication Infrastructure Network.
\end{acknowledgments}

\begin{acknowledgments}
\textbf{Supporting Information Available}: Details on experimental setup, and details on data modeling methods. This material is available free of charge via the Internet at http://pubs.acs.org. 
\end{acknowledgments}

\bibliographystyle{pccp}


\begin{thebibliography}{10}

\bibitem{Mueller2010}
T.~Mueller, F.~Xia, and P.~Avouris, {\em Nat Photon}, 2010, {\bf 4}(5),
  297--301.

\bibitem{Lemme2011}
M.~C. Lemme, F.~H.~L. Koppens, A.~L. Falk, M.~S. Rudner, H.~Park, L.~S.
  Levitov, and C.~M. Marcus, {\em Nano Lett.}, 2011, {\bf 11}(10), 4134--4137.

\bibitem{Bonaccorso2010}
F.~Bonaccorso, Z.~Sun, T.~Hasan, and A.~C. Ferrari, {\em Nat Photon}, 2010,
  {\bf 4}(9), 611--622.

\bibitem{yan2012}
J.~Yan, M.-H. Kim, J.~A. Elle, A.~B. Sushkov, G.~S. Jenkins, H.~M. Milchberg,
  M.~S. Fuhrer, and H.~D. Drew, {\em Nature Nanotechnology}, 2012.

\bibitem{Blake2008}
P.~Blake, P.~D. Brimicombe, R.~R. Nair, T.~J. Booth, D.~Jiang, F.~Schedin,
  L.~A. Ponomarenko, S.~V. Morozov, H.~F. Gleeson, E.~W. Hill, A.~K. Geim, and
  K.~S. Novoselov, {\em Nano Lett.}, 2008, {\bf 8}(6), 1704--1708.

\bibitem{Fong2012}
K.~C. Fong and K.~C. Schwab, {\em Physical Review X}, 2012, {\bf 2}(3), 031006.

\bibitem{Bistritzer2009}
R.~Bistritzer and A.~H. MacDonald, {\em Physical Review Letters}, 2009, {\bf
  102}(20), 206410.

\bibitem{Kubakaddi2009}
S.~S. Kubakaddi, {\em Physical Review B}, {\bf 79}(7), 075417.

\bibitem{Dawlaty2008}
J.~M. Dawlaty, S.~Shivaraman, M.~Chandrashekhar, F.~Rana, and M.~G. Spencer,
  {\em Applied Physics Letters}, 2008, {\bf 92}(4), 042116.

\bibitem{Winnerl2011}
S.~Winnerl, M.~Orlita, P.~Plochocka, P.~Kossacki, M.~Potemski, T.~Winzer,
  E.~Malic, A.~Knorr, M.~Sprinkle, C.~Berger, W.~A. de~Heer, H.~Schneider, and
  M.~Helm, {\em Physical Review Letters}, 2011, {\bf 107}(23), 237401.

\bibitem{Breusing2009}
M.~Breusing, C.~Ropers, and T.~Elsaesser, {\em Physical Review Letters}, 2009,
  {\bf 102}(8), 086809.

\bibitem{Wang2010}
H.~Wang, J.~H. Strait, P.~A. George, S.~Shivaraman, V.~B. Shields,
  M.~Chandrashekhar, J.~Hwang, F.~Rana, M.~G. Spencer, C.~S. Ruiz-Vargas, and
  J.~Park, {\em Applied Physics Letters}, 2010, {\bf 96}, 081917.

\bibitem{Rana2011}
F.~Rana, J.~H. Strait, H.~Wang, and C.~Manolatou, {\em Physical Review B},
  2011, {\bf 84}(4), 045437.

\bibitem{Song2011a}
J.~C.~W. Song, M.~Y. Reizer, and L.~S. Levitov, {\em Physical Review Letters},
  2012, {\bf 109}(10), 106602.

\bibitem{Graham2013}
M.~W. Graham, S.-F. Shi, D.~C. Ralph, J.~Park, and P.~L. McEuen, {\em Nature
  Physics}, 2013, {\bf 9}(2), 103--108.

\bibitem{Betz2013}
A.~C. Betz, S.~H. Jhang, E.~Pallecchi, R.~Ferreira, G.~F\`{e}ve, J.-M. Berroir,
  and B.~Pla\c{c}ais, {\em Nature Physics}, 2013, {\bf 9}(2), 109--112.

\bibitem{Huang2011}
L.~Huang, B.~Gao, G.~Hartland, M.~Kelly, and H.~Xing, {\em Surface Science},
  2011, {\bf 605}(17-18), 1657--1661.

\bibitem{Malard2013}
L.~M. Malard, K.~{Fai Mak}, A.~H. {Castro Neto}, N.~M.~R. Peres, and T.~F.
  Heinz, {\em New Journal of Physics}, 2013, {\bf 15}(1), 015009.

\bibitem{Sun2008}
D.~Sun, Z.-K. Wu, C.~Divin, X.~Li, C.~Berger, W.~A. de~Heer, P.~N. First, and
  T.~B. Norris, October , 2008, {\bf 101}(15), 157402.

\bibitem{Limmer2011}
T.~Limmer, A.~J. Houtepen, A.~Niggebaum, R.~Tautz, and E.~Da~Como, {\em Applied
  Physics Letters}, 2011, {\bf 99}(10), 103104--103104--3.

\bibitem{CastroNeto2009}
A.~H. {Castro Neto}, F.~Guinea, N.~M.~R. Peres, K.~S. Novoselov, and A.~K.
  Geim, {\em Reviews of Modern Physics}, 2009, {\bf 81}(1), 109--162.

\bibitem{Kampfrath2005}
T.~Kampfrath, L.~Perfetti, F.~Schapper, C.~Frischkorn, and M.~Wolf, {\em
  Physical Review Letters}, 2005, {\bf 95}(18), 187403.

\bibitem{tielrooij2013}
K.~J. Tielrooij, J.~C.~W. Song, S.~A. Jensen, A.~Centeno, A.~Pesquera,
  A.~Zurutuza~Elorza, M.~Bonn, L.~S. Levitov, and F.~H.~L. Koppens, April ,
  2013, {\bf 9}(4), 248--252.

\bibitem{Kang2010}
K.~Kang, D.~Abdula, D.~G. Cahill, and M.~Shim, {\em Physical Review B}, 2010,
  {\bf 81}(16), 165405.

\bibitem{Wu2011}
R.~Wu, Y.~Zhang, S.~Yan, F.~Bian, W.~Wang, X.~Bai, X.~Lu, J.~Zhao, and E.~Wang,
  {\em Nano Lett.}, 2011, {\bf 11}(12), 5159--5164.

\bibitem{Xu2009}
X.~Xu, N.~M. Gabor, J.~S. Alden, A.~M. van~der Zande, and P.~L. McEuen, {\em
  Nano Lett.}, 2009, {\bf 10}(2), 562--566.

\bibitem{Gabor2011}
N.~M. Gabor, J.~C.~W. Song, Q.~Ma, N.~L. Nair, T.~Taychatanapat, K.~Watanabe,
  T.~Taniguchi, L.~S. Levitov, and P.~Jarillo-Herrero, {\em Science}, 2011,
  {\bf 334}(6056), 648 --652.

\bibitem{Strait2011}
J.~H. Strait, H.~Wang, S.~Shivaraman, V.~Shields, M.~Spencer, and F.~Rana, {\em
  Nano Lett.}, 2011, {\bf 11}(11), 4902--4906.

\bibitem{Winnerl2012}
S.~Winnerl, F.~Göttfert, M.~Mittendorff, H.~Schneider, M.~Helm, T.~Winzer,
  E.~Malic, A.~Knorr, M.~Orlita, M.~Potemski, M.~Sprinkle, C.~Berger, and
  W.~A.~d. Heer, {\em Journal of Physics: Condensed Matter}, 2013, {\bf 25}(5),
  054202.

\bibitem{Horng2011}
J.~Horng, C.-F. Chen, B.~Geng, C.~Girit, Y.~Zhang, Z.~Hao, H.~A. Bechtel,
  M.~Martin, A.~Zettl, M.~F. Crommie, Y.~R. Shen, and F.~Wang, {\em Physical
  Review B}, 2011, {\bf 83}(16), 165113.

\bibitem{chen2012}
W.~Chen and A.~A. Clerk, {\em Physical Review B}, 2012, {\bf 86}(12), 125443.

\bibitem{Tse2009}
W.-K. Tse and S.~{Das Sarma}, {\em Physical Review B}, 2009, {\bf 79}(23),
  235406.

\bibitem{Du2008}
X.~Du, I.~Skachko, A.~Barker, and E.~Y. Andrei, {\em Nature Nanotechnology},
  2008, {\bf 3}(8), 491--495.

\bibitem{Bolotin2008}
K.~I. Bolotin, K.~J. Sikes, J.~Hone, H.~L. Stormer, and P.~Kim, {\em Physical
  Review Letters}, 2008, {\bf 101}(9), 096802.

\bibitem{Adam}
S.~Adam and S.~D. Sarma, {\em Solid State Communications}, 2008, {\bf
  146}(9–10), 356 -- 360.

\end{thebibliography}

\end{document}